# Switchable ferroelectric bias promoted by boosting square-tensile-strain

Jun Han Lee[1], Nguyen Xuan Duong[2], Min-Hyoung Jung[3], Hyun-Jae Lee[4], Ahyoung Kim[5], Youngki Yeo[6], Junhyung Kim[7], Gye-Hyeon Kim[1], Byeong-Gwan Cho[8], Jaegyu Kim[6], Furqan Ul Hassan Naqvi[9], Jeehoon Kim[10], Chang Won Ahn[2], Young-Min Kim[3], Tae Kwon Song[11], Jae-Hyeon Ko[9], Tae-Yeong Koo[8], Changhee Sohn[1], Kibog Park[1,7], Chan-Ho Yang[6], Sang Mo Yang[5], Jun Hee Lee[4], Hu Young Jeong[12], Tae Heon Kim[2,*], and Yoon Seok Oh[1,*]

[1]*Department of Physics, Ulsan National Institute of Science and Technology (UNIST), Ulsan 44919, Republic of Korea*

[2]*Department of Physics and Energy Harvest-Storage Research Center (EHSRC), University of Ulsan, Ulsan 44610, Republic of Korea*

[3]*Department of Energy Science, Sungkyunkwan University, Suwon, 16419 Republic of Korea*

[4]*School of Energy and Chemical Engineering, Ulsan National Institute of Science and Technology (UNIST), Ulsan, 44919, Republic of Korea*

[5]*Department of Physics, Sogang University, Seoul 04107, Republic of Korea*

[6]*Department of Physics & Center for Lattice Defectronics, Korea Advanced Institute of Science and Technology (KAIST), Daejeon 34141, Republic of Korea*

[7]*Department of Electrical Engineering, Ulsan National Institute of Science and Technology, Ulsan 44919, Republic of Korea*

[8]*Pohang Accelerator Laboratory, Pohang University of Science and Technology (POSTECH), Pohang, Gyeongbuk 790-784, Republic of Korea*

[9]*School of Nano Convergence Technology, Nano Convergence Technology Center, Hallym University, Chuncheon 24252, Republic of Korea*

[10]*Department of Physics, Pohang University of Science and Technology (POSTECH), Pohang, Gyeongbuk 790-784, Republic of Korea*

[11]*Department of Materials Convergence and System Engineering, Changwon National University, Changwon, Gyeongnam, 51140, Republic of Korea*

[12]*UNIST Central Research Facilities, Ulsan National Institute of Science and Technology, Ulsan 44919, Republic of Korea*

* Corresponding author. Email: thkim79@ulsan.ac.kr; ysoh@unist.ac.kr

Interaction between dipoles often emerges intriguing physical phenomena, such as exchange bias in the magnetic heterostructures and magnetoelectric effect in multiferroics, which lead to advances in multifunctional heterostructures. However, the defect-dipole tends to be considered the undesired to deteriorate the electronic functionality. Here, we report deterministic switching between the ferroelectric and the pinched states by exploiting a new substrate of cubic perovskite, $BaZrO_3$, which boosts square-tensile-strain to $BaTiO_3$ and promotes four-variants in-plane spontaneous polarization with oxygen vacancy creation. First-principles calculations propose a complex of an oxygen vacancy and two $Ti^{3+}$ ions coins a charge-neutral defect-dipole. Cooperative control of the defect-dipole and the spontaneous polarization reveals three types of in-plane polar states characterized by biased/pinched hysteresis loops. Furthermore, we experimentally demonstrate that three electrically controlled polar-ordering states lead to switchable and non-volatile dielectric states for application of non-destructive electro-dielectric memory. This discovery opens a new route to develop functional materials via manipulating defect-dipoles and offers a novel platform to advance heteroepitaxy beyond the prevalent perovskite substrates.

## 1 Introduction

Anisotropic interactions between order parameters have provided innovative pathways for developing functional devices on the peculiar physical properties. For example, anisotropic exchange interaction at the interface between a ferromagnetic and an antiferromagnetic layer generates exchange bias effect (Fig. 1a), which has been a key property to realize magnetic readback heads and magnetic random access memory devices[1, 2]. Antisymmetric exchange interaction, known as Dzyaloshinky-Moriya interaction, between neighboring magnetic moments builds up magnetically induced electric polarization and magnetoelectric effect in multiferroic[3]. For ferroelectrics, however, the anisotropic interactions between electric dipoles have been imitated by spatial electric potential gradient, rather than a cooperative dipolar unit element, resulting in imprint and voltage shift in ferroelectric hysteresis loop $P(E)$ [4-7]. It has been extensively studied that the spatial electric potential gradient originates from various sources, such as charge trapping/migration/injection[8], built-in field attributed to a difference in the work function between the asymmetric top and bottom electrodes[9], strain-gradient-induced flexoelectric field[10], surface polar absorbates[11], and interfacial Schottky effect[12]. Such spatial potential gradient easily appears in the vertical geometry of ferroelectric heterostructures and provokes the instability (e.g., polarization fatigue/relaxation, retention loss, depolarization, and self-polarization) of switchable polarization[13]. Furthermore, the congenital slow dynamics or stationary property circumscribe versatile application and technological advances in functionality of the non-volatile ferroelectric memory devices.

For ferroelectric properties, the non-centric charge distribution of the oxygen vacancy $V_O$ produces a defect-dipole $P_d$, which has been known to deteriorate ferroelectricity or to enhance piezoelectricity[14]. Here, we take advantage of anisotropic interaction between $P_d$ and spontaneous polarization $P_{FE}$, and demonstrate to reversibly manipulate the direction of internal bias-field $E_{bias}$ in the ferroelectric hysteresis loop similar to the exchange bias effect in magnetic heterostructures (Fig. 1a)[1, 2]. In order to prohibit the spatially asymmetric electric potential gradient and construct isotropic geometry of a device, we design in-plane polarized ferroelectric heterostructures within the horizontal geometry by applying square tensile stress in the (001) plane of ferroelectric $BaTiO_3$[15]. Our first-principles calculations based on density functional theory (DFT) predict that a $V_O$ produces a defect-dipole complex in the spontaneous polar phase of $BaTiO_3$ under the square-tensile-strain (Fig. 1b)[16]. Fig. 1b illustrates the theoretical results of local $Ti^{3+}$ and $O^{2-}$ ion displacements when a $V_O$ is located at a vertex of two corner-sharing $TiO_6$ octahedra along the [010]-axis and the $P_{FE}$ is oriented along the [100]-direction in square-tensile-strained (001) plane of $BaTiO_3$. We find that the canted displacement of $Ti^{3+}$ ions for a $V_O$ at a vertex of the [010]-axis produces a finite $P_d$ along the opposite direction of $P_{FE}$//[100], exhibiting the lowest total energy of the system (Fig. 1b and Fig. S1e). Thus, the canted $Ti^{3+}$-$V_O$-$Ti^{3+}$ complex, consisting of two $Ti^{3+}$ ions and a $V_O$, builds up $P_d$ in the opposite direction of $P_{FE}$, where the $P_d$ can act as a source of a local internal field in polarization reversal microscopically. For the as-grown pristine state, $P_{FE}$ could develop 90° domain walls with four polarization variants in the square strained (001) plane to minimize electrostatic energy[15, 17] and induce an energetically favored antiparallel $P_d$ to the $P_{FE}$ (Fig. 1a). The 90° domains and antiparallel $P_d$-$P_{FE}$ arrangement would result in the pinched hysteresis loop for the pristine state[14]. On the other hand, external application of the electric field under sufficiently high thermal energy for the $V_O$ to migrate in the lattice[18] could induce a single domain and manipulate the direction of the $P_d$, characterized by biased ferroelectric hysteresis loop with the orientational dependent $E_{bias}$. We successfully demonstrate deterministic control of the $E_{bias}$ in $P(E)$, and thereby, three types of in-plane polar configurations are achievable in the remanent state at room temperature.

## 2 Results and Discussion

In order to induce square-tensile-strain to the BaTiO$_3$ film, we develop a new substrate of cubic perovskite oxide BaZrO$_3$ with a large lattice constant of 4.189 Å (Fig. 2a and Fig. S2) [16]. The BaTiO$_3$ epitaxy films on the BaZrO$_3$ substrates are grown on the (001) surface by pulsed laser deposition (PLD)[16]. Annular dark-field (ADF) scanning transmission electron microscopy (STEM) images (Fig. 2b) represents that the BaTiO$_3$ film is epitaxially grown on the BaZrO$_3$ substrate. The reciprocal space maps (RSM) of {204} Bragg reflections of the BaTiO$_3$ film (Fig. 2C and Fig. S8) determine that the out-of-plane $c$//[001], in-plane $a$//[100], and $b$//[010] lattice constants of the BaTiO$_3$ film are $c$=4.009(7) Å and $a$=$b$=4.039(3) Å, respectively. The azimuthal angle ($\phi$)-independent $|Q_x|$(or $|Q_y|$) and $Q_z$ values for the {204} reflections of the BaTiO$_3$ film indicate that the lattice of the BaTiO$_3$ film maintains 4-fold symmetry in the (001) plane[16]. In addition, Raman spectra of the BaTiO$_3$ film (Fig. S9) clearly show two predominant peaks at 301 and 526 cm$^{-1}$ correspond to representative Raman spectra of the tetragonal BaTiO$_3$ bulk single-crystal[19]. Thus, both experiments indicate that the in-plane square tensile stress of the BaZrO$_3$ substrate leads to a $c/a$=0.99<1 tetragonal lattice distortion of the epitaxial BaTiO$_3$ film rather than an orthorhombic[20] or a monoclinic[21] distortion, at least in these measurements scale. In the electron energy-loss spectroscopy (EELS) spectra (Fig. 2d), we observe the decreased near-edge fine structure of the BaTiO$_3$ film (Fig. 2d) in comparison with that of the bulk BaTiO$_3$ single-crystal. It reveals that the BaTiO$_3$ film incorporates more V$_{OS}$ than the bulk BaTiO$_3$ single crystal.

The *ab initio*-based molecular dynamic simulations in a box of 32×32×32 unit cells expected that the square-tensile-strain on BaTiO$_3$ stabilizes a multi-domain ferroelectric phase with electric polarization along <100>, in which the elastic boundary condition through the epitaxial constraint of the square-tensile-strain favors the tetragonal phase[15]. We performed in-plane piezoresponse force microscopy (IP-PFM) measurements of a pristine BaTiO$_3$ film (Fig. 2e, 2f, and Fig. S12)[16], which reveal four-variants polar domains with irregular distribution and meandering boundaries in nanoscale. For the archetypal ferroelectric domain structures of BaTiO$_3$, the crystallographic symmetry and saved electrostatic energy at the domain walls favor the stripe patterns of 90° domain walls[15, 17]. On the other hand, in epitaxial ferroelectric films, the substrate surface symmetry constrains elastic variants and strongly affects the domain structures. For example, four in-plane ferroelectric variants of BiFeO$_3$ films are constrained on SrTiO$_3$ (110) substrates[22]. In the epitaxial BaTiO$_3$ film on the BaZrO$_3$ substrates, spontaneous in-plane polarization breaks the 4-fold symmetry in the *ab*-plane and should allow orthorhombic 2-fold symmetry. Nevertheless, the 4-fold symmetric square lattice of the BaZrO$_3$ (001) surface could compel the four-variants in-plane polar domains. In the macroscopic scale, the averaged structural symmetry of the orthorhombic nano-domains with four-variants would appear to be 4-fold tetragonal, which is in agreement with our results observed in the XRD and Raman analyses. From this experimental evidence of a *c*/*a*<1 tetragonal lattice distortion (characterized by macroscopic XRD and Raman measurements) and four-variants polar domains (microscopically observed in the IP-PFM measurements), we conclude that the square tensile stress of the BaZrO$_3$ substrate produces a tetragonally distorted BaTiO$_3$ film with four-variants orthorhombic (*Pmm*2) nano-domains, where spontaneous polarization breaks the 4-fold symmetry in the *ab*-plane of *a*=*b*>*c*.

Fig. 3a-3c represent experimental results of $P(E)$ measurements of a BaTiO$_3$ film with the interdigital electrodes along the [100]- and [110]-directions (inset of Fig. S14a)[16]. The pristine state of the BaTiO$_3$ film shows pinched $P(E)$, where the saturated electric polarizations $P_s$ are $P_{s,[100]}$=11.5 µC/cm$^2$ for [100] and $P_{s,[110]}$=10.2 µC/cm$^2$ for [110], as observed in the aged ferroelectrics. Under sufficiently high thermal energy for V$_O$ to migrate in the lattice, the $P_d$ is oriented along with directions of the applied electric field or the $P_{FE}$[18]. We designed a poling process[5] to align and switch the $P_d$ direction in a square-tensile-strained BaTiO$_3$ film (Fig. S13a). The specimen was cooled down under a constant applied electric field $E_{pole}$ along the [100]- or [110]-direction from $T$=120 °C, and then, after $E_{pole}$ was turned off at room temperature ($T$=27 °C), $P(E)$ was measured. The positively poled state reveals distinct negative-biased ferroelectric $P(E)$ (Fig. 3b) with bias-fields $E_{bias,[100]}$=-33.3 kV/cm for [100] and $E_{bias,[110]}$=-43.3 kV/cm for [110], where the anisotropy of $P_s$ ($P_{s,[100]}$=12.9 µC/cm$^2$ and $P_{s,[110]}$=9.1 µC/cm$^2$) increases to $P_{s,[100]}/P_{s,[110]} \sim \sqrt{2}$. Negative poling switches to positive-biased ferroelectric $P(E)$ while maintaining the anisotropy and magnitude of $P_s$ (Fig. 3c). The $E_{bias}$ is reversibly manipulated by thermal treatment and the subsequent electrical poling. In addition, zero-field cooling (ZFC)

restores the pinched $P(E)$ and preserves the anisotropic $P_s$ (Fig. S14). Thus, three types of in-plane polar states with distinct $P(E)$ are reproducibly controlled by the electric field cooling process[16].

In addition, the $E_{bias}$ in the $P(E)$ measurements corresponds to an internally induced electric field by the $P_d$. Contrary to the imprint due to the spatial potential gradients[5, 7], the $E_{bias,[110]}$ and $E_{bias,[100]}$ of the BaTiO$_3$ film exhibit the apparent anisotropy of $E_{bias,[110]}/E_{bias,[100]}=1.3\sim\sqrt{2}$ (Fig. 3b and 3c). A combination of V$_O$ existence and anisotropic $E_{bias}$ provides evidence that the $E_{bias}$ of the BaTiO$_3$ film originates from an orientational dependent $P_d$ of the Ti$^{3+}$-V$_O$-Ti$^{3+}$ complex in the lattice. Epitaxial strain, lattice expansion, the energetics of the PLD growth, and non-stoichiometry related effects could yield cations (Ba$^{2+}$, Ti$^{4+}$) and oxygen vacancies, which form the $P_d$ for the complex oxide thin films [6, 23, 24]. Rather than other vacancies of the cations, the V$_O$ requires the order of magnitude lower thermodynamic energy to get mobility[23]. The poling temperature of $T=120$ °C for the $P_d$ is within that of the V$_O$ for our poling process[23]. Therefore, the $P_d$ associated with switchable $E_{bias}$ for the three types of polar states is attributed to the V$_O$ rather than the cation vacancies.

Adopting the canted Ti$^{3+}$-V$_O$-Ti$^{3+}$ complex model for $P_d$, we illustrate how $P_{FE}$, $P_d$, and V$_O$ cooperate in the $P(E)$ measurements (Fig. 3, right). At $E=0$ kV/cm of the pristine state (Fig. 3a), $P_{FE}$ develops 90° domain walls to minimize electrostatic energy[15, 17] and induces an energetically favored antiparallel $P_d$ to the $P_{FE}$. Applying external $E$ to the pristine aligns the $P_{FE}$ to the direction of the $E$ but not the $P_d$ because of insufficient thermal energy at room temperature [18]. The $P_d$ plays a role in restoring force to reverse $P_{FE}$ and results in pinched $P(E)$[14]. Increasing the temperature reduces the energy barrier for V$_O$ migration[25]. At the high temperature, application of the $E_{pole}$ along the [100]-direction aligns both $P_{FE}$ and $P_d$ in parallel along the applied $E_{pole}$ direction, in which V$_O$ at of the [100]-axis migrates to the nearest vertex of the [010]-axis (Fig. 3b, right). So, the poling process leads to a single domain of both $P_{FE}$ and $P_d$ along the [100]-direction. At room temperature, the applied $E$ in the $P(E)$ measurements becomes, albeit sufficiently large for switching $P_{FE}$, insufficient to switch $P_d$. Thus, $P_d$ returns to a role in the imprinted internal electric field (Fig. 3b). As a result, the ferroelectric $P(E)$ of the positively (negatively) poled state is negatively (positively) biased.

An exclusive feature of the ferroelectric switchable bias, including the three types of in-plane polar states, reflects that the application of the external electric field and the pre-poling systematically controls each $P_{FE}$ and $P_d$, and determines the anisotropic interactions between $P_{FE}$ and $P_d$, similar to the switchable exchange bias effect in magnetic heterostructures[2]. We also expect that the cooperation of $P_{FE}$ and $P_d$ gives rise to distinct dielectric responses with respect to the external electric field. Fig. 4a, 4b, and 4c represent DC electric field $E_{DC}$ dependence of the dielectric constant $\varepsilon(E_{DC})$ for the ZFC, positively poled, and negatively poled states, respectively (Fig. S14). For the ZFC state, $\varepsilon(E_{DC})$ shows an archetypal symmetric butterfly shape(Fig. 4a). On the other hand, the positively (Fig. 4b) and the negatively (Fig. 4c) poled states exhibit biased $\varepsilon(E_{DC})$, which result in discrepant values of $\varepsilon(E_{DC}=0)=1258$ and 1324 at the zero $E_{DC}$. Our schematic picture of $P_{FE}$ and $P_d$ domains reveals that high and low values of $\varepsilon(E_{DC}=0)$ are determined by antiparallel and parallel alignments between $P_{FE}$ and $P_d$, respectively. Remarkably, as shown in Fig. 4d, the high and low dielectric states of $\Delta\varepsilon$~6.5 % in the initial switching are alternatively reproducible with successive bipolar pulses of $E_{DC}$ for the positively and the negatively poled states. The repeated switching between the two dielectric states accompanies dielectric relaxation as a function of $A\exp(-t/\tau)+\varepsilon$, where $A$, $\tau$, and $\varepsilon$ for the high(low) dielectric state are $A$~25 (-12), $\tau$~198 (83) seconds, and $\varepsilon$~1330 (1300), respectively, and reaches $\Delta\varepsilon$~2.3 %. The ZFC state recovers the unpoled/zero-biased state of $\varepsilon(E_{DC}=0)=1316$. This demonstrates that the electrically controlled antiparallel/parallel arrangements of $P_{FE}$ and $P_d$ realize non-volatile $\varepsilon(E_{DC}=0)$ states (Fig. 4e).

## 3 Conclusion

This study demonstrates the switchable three polar states in a single layer of in-plane ferroelectric heterostructures by promoting defect-dipole and four-variants orthorhombic ($Pmm2$) nano-domains, beyond the binary polar states[4-7, 26]. The development of four-variants polar domains of the $BaTiO_3$ on the $BaZrO_3$ substrate presents that the large isotropic surface lattice can spawn a novel ground state and physical phenomena in other inaccessible heterostructures, such as two-dimensional topological phases of honeycomb superlattices on 6-fold symmetric (111) surface[27]. The $BaZrO_3$ substrate will be harnessed as a new platform for artificial design to a conceptual material system via heteroepitaxy inevitably combined with strain engineering. In addition, the switchable dielectric states inspire that the dielectric constant, rather than the electrical resistivity, can be considered a low-energy-consumption memory information.

## 4 Experimental Section

**BaZrO₃ single-crystal growth, substrate preparations, and physical properties**

BaZrO$_3$ is one of a few cubic perovskite oxides and has a large lattice constant of 4.189 Å (Fig. S2a). The high melting temperature $T$=2690 °C[28] of BaZrO$_3$ and severe evaporation of BaO inhibit single crystal growth under ordinary substrate growth environments, and high-quality growth can be achieved using the optical floating zone method[29] and the induction skull melting method[30]. We successfully grow a 4 cm long BaZrO$_3$ single-crystal of ~4 mm diameter (Fig. S2b) by optimizing the growth condition of the reported optical floating zone method[29]. Polycrystalline BaZrO$_3$ feed rods were prepared as stoichiometric BaO and ZrO$_2$ and mixed, ground, pelletized, and sintered at 1650 °C for 24 hr in air. BaZrO$_3$ single-crystals were grown using the optical floating zone (FZ-T-12000-X-VII-VPO-PC, Crystal System) in a 10 % O$_2$ and 90 % Ar mixed gas environment under a pressure of 5 bar and flow rate of 4 L/min. Anti-clockwise rotation of the feed and seed rods at 20 rpm were carried out. The travelling speeds, optimized to maintain the stable molten zone, were 12.9 and 9.2 mm/hr for the feed and seed rods, respectively. The as-grown single-crystals were annealed at 1650 °C in O$_2$ flow. The grown BaZrO$_3$ single-crystal rod had a diameter of 4–5 mm and length of ~4 cm as shown in Fig. S2b. The sliced BaZrO$_3$ single-crystal disks were polished using an Allied High Tech Multiprep Polishing System and a Pace Technologies GIGA-1200 Vibratory Polisher.

The cylindrical axis of the grown BaZrO$_3$ single-crystal tends oriented along the crystallographic [001] axis (Fig. S2c). Once the (001) cleaved surface is achieved by cleaving perpendicular to the cylindrical axis, the (001) BaZrO$_3$ substrate is prepared by slicing or cleaving another (001) surface and polishing both surfaces (Fig. S2d). The root-mean-square (RMS) surface roughness of the polished (001) surface is as low as 1.58 Å (Fig. S2e). The full width half maximum (FWHM) from the rocking curve (Fig. S2f) is 0.021–0.075°, comparable to that of the prevalent SrTiO$_3$ substrates. BaZrO$_3$ is an ultrawide bandgap semiconductor[31]; hence, optical transmittance spectra (Fig. S2g) and Tauc plots (Fig. S2h and S2i) exhibit direct 4.97 eV and indirect 4.88 eV energy bandgaps. The near-edge defect state, resolved from the Urbach tail in Tauc plots, is negligible in comparison with both the reported and commercial single-crystals[29]. The dielectric constant $\varepsilon$ and loss tan$\delta$ at $f$=20 kHz along the [001] axis are ~55 and ~0.001, respectively, at room temperature (Fig. S2j). As the temperature decreases, $\varepsilon$ gradually increases and a weak dielectric anomaly appears at $T$=150 K with a peak of tan$\delta$ as observed in previous reports[32, 33]. These studies attribute such dielectric anomaly to several reasons, e.g. polaronic relaxation[34], unavoidable impurities[32], and dynamic Jahn-Teller-like octahedra distortion[35]. Still, the origin of this anomaly is unclear and requires further comprehensive investigation. BaZrO$_3$ consists of only diamagnetic, so-called non-magnetic, Ba$^{2+}$, Zr$^{4+}$, and O$^{2-}$ ions; thus, the magnetic susceptibility is negative, $\chi_{dc}$~-55×10$^{-6}$ emu/mole, and nearly temperature independent.

**Film growth**

BaTiO$_3$ films were epitaxially fabricated on BaZrO$_3$ (001) substrates using the pulsed laser deposition (PLD) method. The film thickness of all BaTiO$_3$ films is 60–70 nm. The epitaxy and the thickness of the as-grown BaTiO$_3$ (001) films were characterized by a lab-based x-ray diffractometer (D8 Advance, Bruker). A pulsed excimer laser (KrF, wavelength of 248 nm) was irradiated into a BaTiO$_3$ ceramic target to generate a plasma plume for film growth. The laser fluence for the PLD growth was about 1.1 J/cm$^2$. Before the actual film deposition, the surface of the ceramic target was pre-ablated with a pulsed laser for a particular duration. BaTiO$_3$ films were deposited at 650 °C under an oxygen partial pressure of 20 mTorr. Subsequently, *in situ* post-annealing was performed at 630 °C for 1 hr under the oxygen environment of 100 Torr. In addition, all BaTiO$_3$ films are *ex-situ* annealed at 600 °C for 3 hr in an oxygen atmosphere under the ambient pressure in the tube furnace. The specimen's state after the film growth and the *ex-situ* annealing is called pristine state.

**Crystallographic structure of BaTiO$_3$ Film on the BaZrO$_3$ substrate**

The full width half maximum (FWHM) in (002) rocking curves of the *ex-situ* annealed BaTiO$_3$ films (Fig. S3) are 0.15–0.40°. ADF-STEM, the temperature dependence of XRD, EELS, $P(E)$, and electro-dielectric memory effect were examined on film-a. Contact resonance of the piezo-response force microscopy were measured on film-b. XRD, RSM, Raman scattering, and IP-PFM were performed on film-c. The 66 nm film thickness of film-a was verified by the cross-sectional ADF-STEM images (Fig. S4a). Even though the large lattice mismatch of 4.49 % between BaTiO$_3$ and BaZrO$_3$ leads to the edge dislocation at the interface between the BaTiO$_3$ film and the BaZrO$_3$ substrate (Fig. S4b), we successfully grew the epitaxial BaTiO$_3$ (001) films, in which FWHM are 0.15–0.40°. Due to the large nominal misfit strain of 4.49 % between BaTiO$_3$ and BaZrO$_3$, the BaTiO$_3$ lattice constants are observed to rapidly relax, when the BaTiO$_3$ film layer is away from the BaTiO$_3$/BaZrO$_3$ interface (Fig. S5).

$\theta$-$2\theta$ x-ray diffraction (XRD) shows clear (00*l*) Bragg reflections of the BaTiO$_3$ film and BaZrO$_3$ substrate (Fig. S6). A $\phi$-scan of {101} Bragg peaks of the BaTiO$_3$ film and BaZrO$_3$ substrate (Fig. S7) present that the BaTiO$_3$ film is epitaxially grown on the BaZrO$_3$ substrate. From the reciprocal space maps (RSM) of {204} Bragg reflections of the BaTiO$_3$ film (Fig. S8), we determine that the out-of-plane *c*, in-plane *a*, and *b* lattice constants of the BaTiO$_3$ film are *c*=4.009(7) Å and *a*=*b*=4.039(3) Å, respectively. A tensile strain $\eta$=+0.98 % ($\eta$=100×(*a*/$a_{pc,0}$-1), where $a_{pc,0}$ is the cube root of volume of bulk tetragonal BaTiO$_3$ at *T*=20 °C; $a_{pc,0}$=4.000(4) Å[36]) is applied to the 66 nm thick BaTiO$_3$ film.

Raman spectra of the BaTiO$_3$ film (Fig. S9) clearly show two predominant peaks at 301, and 526 cm$^{-1}$ correspond to representative Raman spectra of the tetragonal BaTiO$_3$ bulk single-crystal[19]. Thus, both experiments indicate that the square tensile stress of the BaZrO$_3$ substrate leads to a *c*/*a*=0.99<1 tetragonal lattice distortion of the epitaxial BaTiO$_3$ film rather than an orthorhombic[20] or a monoclinic[21] distortion, at least in these measurements scale.

The change in slope of the temperature dependence of the lattice constant has often been referred to as evidence of a ferroelectric phase transition[20, 37]. The temperature dependence of the in-plane lattice constant of our square-tensile-strained BaTiO$_3$ film on the BaZrO$_3$ substrate exhibits a change in slope at ~415 °C (Fig. S10), close to the predicted $T_c$ from *ab initio* calculations[15] and higher than that on MgAl$_2$O$_4$ substrate[20].

**PFM measurements**

The space group *Pmm*2 has 2mm symmetry, where the 2-fold axis is parallel to the polar [100] axis. The symmetry allows the following piezoelectric tensor.

$$d_{ij} = \begin{pmatrix} d_{11} & d_{12} & d_{13} & 0 & 0 & 0 \\ 0 & 0 & 0 & 0 & 0 & d_{26} \\ 0 & 0 & 0 & 0 & d_{25} & 0 \end{pmatrix}$$

where index *i*, *j* = 1, 2, 3 correspond to [100], [010], [001] axes. The conventional geometry of the IP-PFM measurement probes the shear piezoelectric response $d_{25}$ of the *Pmm*2 orthorhombic BaTiO$_3$ film.

We conducted in-plane PFM measurement over an area of 1 µm × 1 µm with a condition of 256 lines using a commercial scanning probe microscope (Asylum Research MFP 3D Infinity, Oxford

Instruments). Conductive Pt-coated silicon tips (MikroMasch, HQ:NSC35/Pt) were used. Tip scanning speed was 2 μm/s and an AC voltage of 3 V was applied to the tip during PFM scans. The longest cantilever tip with 130 μm was selected to avoid large loading forces imposed on the sample. To enhance the signal, we measured PFM images in near resonance conditions at 1.29 MHz. The reproducibility of the in-plane PFM images was verified by correcting the signal offset and taking several consecutive measurements on the same area.

Contact resonance of the piezo-response force microscopy (PFM) in Fig. S11 were performed by an atomic force microscope (NX10, Park Systems) with a lock-in amplifier (HF2LI, Zurich Instruments). We performed IP-PFM measurements through the lateral oscillation of a non-conductive cantilever while an AC voltage $V_{ac}$ of ±5 V was applied to a pair of interdigital electrodes along the [110] or [100] directions[38]. To obtain a high signal-to-noise ratio, dual-amplitude-resonance-tracking PFM mode was used. Note that the $BaTiO_3$ film has no bottom electrodes and non-conductive cantilevers (PPP-FMR, Nanosensors) were used. The applied $V_{ac}$ of 5 V between 5 μm gaps of the interdigital electrode's fingers corresponds to the electric field of 10 kV/cm, 22 % of the coercive field in the $P(E)$ of the electrically poled state. The first harmonic contact resonance frequency of the in-plane PFM signals was approximately 770 kHz (Fig. S11). The in-plane piezo-response, measured by the PFM, is a factor of ten larger than the out-of-plane piezo-response (Fig. S11). These results indicate that the predominant piezo-response of our epitaxial $BaTiO_3$ film on the $BaZrO_3$ substrate arises from in-plane components of the film rather than out-of-plane direction (the [001] direction).

**Acknowledgments**

We thank C. B. Eom and S.-W. Cheong for constructive comments on the manuscript. **Funding:** This work was supported by the Basic Science Research Programs through the National Research Foundation of Korea (NRF) (NRF-2020R1A2C1009537). N.X.D., C.W.A. and T.H.K. acknowledge support from the Priority Research Centers Program through the National Research Foundation of Korea (NRF) funded by the Ministry of Education (Grant No. NRF-2019R1A6A1A11053838). H.Y.J. acknowledges support from the Creative Materials Discovery Program through the National Research Foundation of Korea (NRF-2016M3D1A1900035). Initial preliminary work on polishing substrate by Y.S.O. at Rutgers was supported by the visitor program at the center for Quantum Materials Synthesis (cQMS), funded by the Gordon and Betty Moore Foundation's EPiQS initiative through grant GBMF10104, and by Rutgers University.

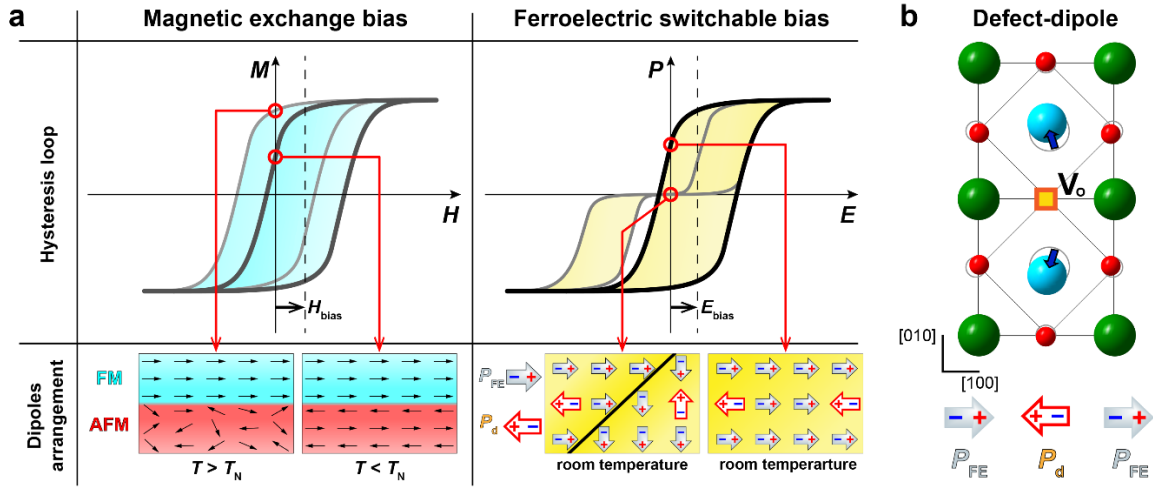

**Fig. 1. Switchable ferroelectric bias in hysteresis loop and theoretical calculations for defect-dipole of a $Ti^{3+}$-$V_O$-$Ti^{3+}$ complex. a**, Anisotropic interactions and characteristic hysteresis loops of magnetic heterostructures and ferroelectrics. Exchange interaction at an interface between ferromagnetic and antiferromagnetic layers leads to $H_{bias}$ biased hysteresis loop of the magnetic heterostructures. For ferroelectrics, defect-dipole $P_d$ in the lattice can play a role in the bias-field $E_{bias}$. Cooperation of the $P_d$, spontaneous polarization $P_{FE}$, and domain arrangement enables deterministic control of ferroelectric bias, characterized by either pinched or biased hysteresis loop. **b**, Density functional theory (DFT) calculations simulate local $Ti^{3+}$ and $O^{2-}$ ion displacements of defect-dipole when the $P_{FE}$ is oriented along the [100]-direction and an oxygen vacancy $V_O$ is located at a vertex of the [010]-axis in a $TiO_6$ octahedron[16]. The green, blue, and red spheres represent barium ($Ba^{2+}$), trivalent titanium ($Ti^{3+}$), and oxygen ($O^{2-}$) ions, respectively. A yellow square indicates a $V_O$. The open grey circles show the ions' initial positions without the $V_O$. The dark blue arrows illustrate the $Ti^{3+}$ ion displacements from the initial positions. Two $Ti^{3+}$ ions neighboring the $V_O$ are displaced away from the $V_O$ so that the $V_O$ leads to canted displacement of the $Ti^{3+}$ ions and finite $P_d$ of the canted $Ti^{3+}$-$V_O$-$Ti^{3+}$ complex.

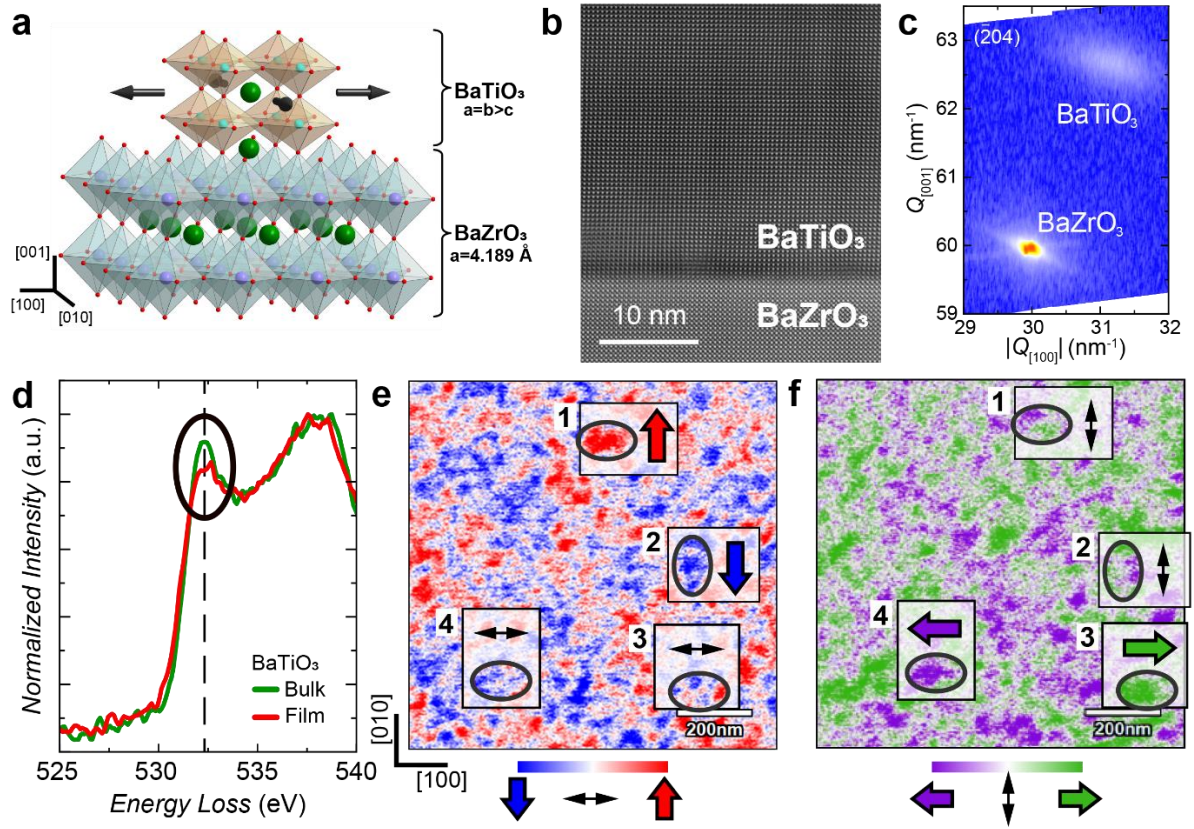

**Fig. 2.** $c/a<1$ **tetragonally distorted BaTiO$_3$ film on the cubic BaZrO$_3$ substrate and four-variants in-plane polar domains. a,** Schematic of a $c/a<1$ tetragonal unit cell of square-tensile-strained BaTiO$_3$ film on the cubic perovskite BaZrO$_3$ substrate. **b,** Cross-sectional ADF-STEM image of the BaTiO$_3$ film and BaZrO$_3$ substrate. **c,** Reciprocal space maps (RSMs) at ($\bar{2}$04) reflections for the BaTiO$_3$ film and BaZrO$_3$ substrate. Here, the [001]-direction corresponds to out of the sample surface plane. Strong and sharp BaZrO$_3$ substrate peaks exist near $Q_{[001]}$=59.96 nm$^{-1}$ and $|Q_{[100]}|$=29.98 nm$^{-1}$, and weak BaTiO$_3$ film peaks are observed near $Q_{[001]}$=62.68 nm$^{-1}$ and $|Q_{[100]}|$=31.11 nm$^{-1}$, which indicate $c$=4.009(7) Å and $a$=$b$=4.039(3) Å, respectively. **d,** Electron energy-loss spectroscopy (EELS) results of the O $K$-edge taken from the 66 nm thick BaTiO$_3$ film (red) and the BaTiO$_3$ bulk single-crystal (green). The EELS spectra of the BaTiO$_3$ film show a decreased near-edge fine-structure at 532.3 eV, which indicates the existence of oxygen vacancies V$_O$s in the BaTiO$_3$ film. The dashed line guides the normalized intensity at energy losses of 532.3

eV. **e,f**, In-plane piezoresponse force microscope (IP-PFM) image simultaneously taken by a tip of which the cantilever axis orients along (**e**) [100]- and (**f**) [010]-directions. The colored solid arrows imply the direction of the in-plane polarization.

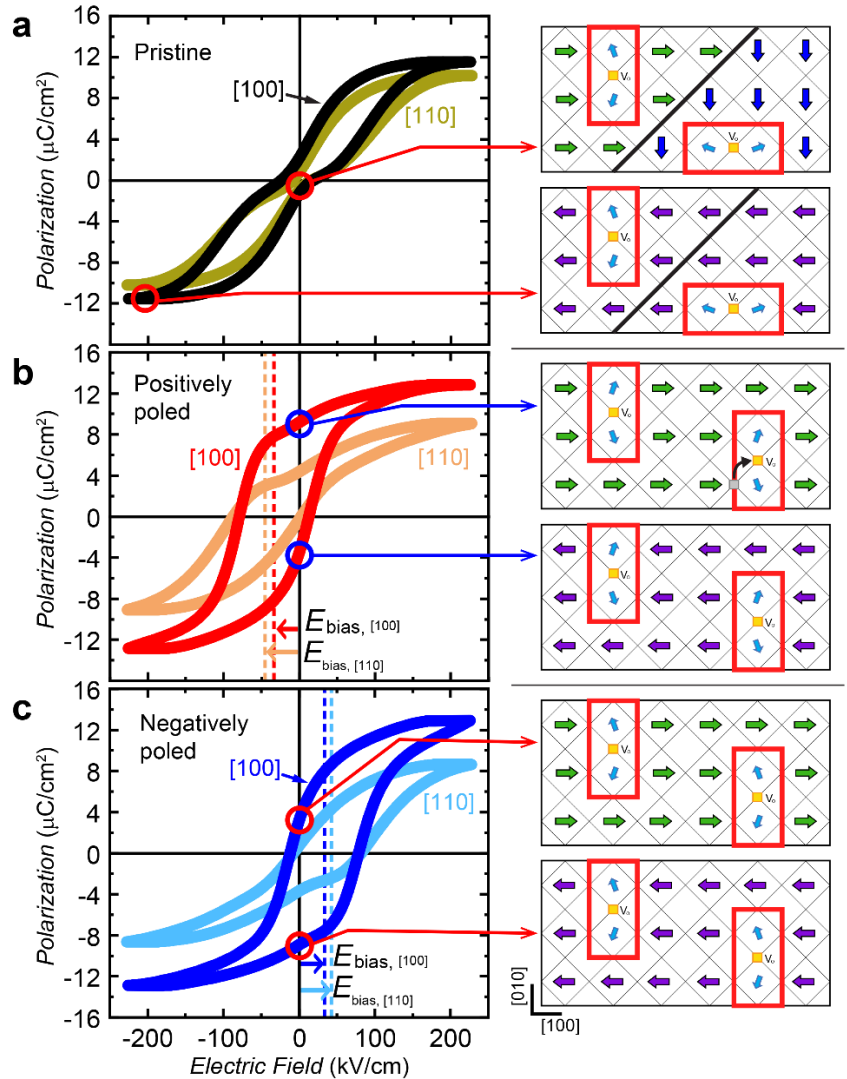

**Fig. 3. Three types of in-plane polar states and switchable ferroelectric bias.** $P(E)$ hysteresis loop (left) and schematic polar domains (right) of (**a**) the pristine, (**b**) the positively poled, (**c**) the negatively poled BaTiO$_3$ film. The $P(E)$ hysteresis loops were measured along the [100]- and [110]-directions at $T$=27 °C. In the schematic of the polar domain, the colored solid arrows depict the $P_{FE}$, and the red rectangles are Ti$^{3+}$-V$_O$-Ti$^{3+}$ complex of a defect-dipole. **a,** The pristine BaTiO$_3$ film favors the head-to-tail 90° domain wall of the $P_{FE}$ parallel to the {110} plane for saving electrostatic energy. The finite $P_d$s are aligned opposite to $P_{FE}$ at $E = 0$ kV/cm of the pristine state. **b,c,** Poling dependent bias-field in the ferroelectric $P(E)$ along the [100] (dark colors) and [110]

(light colors) directions. The positive and negative polings produce negatively and positively biased ferroelectric $P(E)$, respectively, where the bias-fields $E_{bias,[100]}=\mp33.3$ kV/cm along the [100] and $E_{bias,[110]}=\mp43.3$ kV/cm along the [110]. The positive poling along the [100]-direction at high temperature aligns both $P_d$ and $P_{FE}$ along the [100]-direction even through $V_O$ hopping, in which the $V_O$ at a vertex of the [100]-direction migrates to the neighboring vertex of the [010]-direction for aligning $P_d$ parallel to the poling voltage along the [100]-direction. The black arrow represents the $V_O$ hopping.

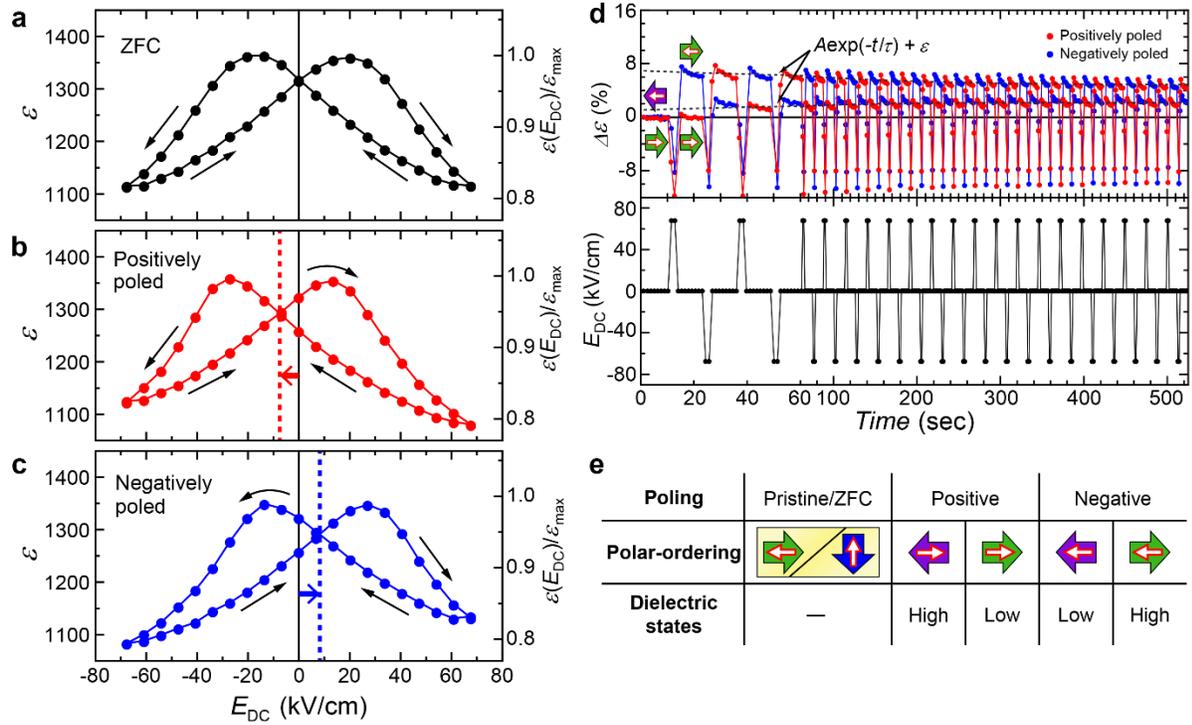

**Fig. 4. Switchable dielectric memory effect. a,b,c,** DC electric field $E_{DC}$ dependent dielectric constant $\varepsilon$ (left axes) and $\varepsilon(E_{DC})/\varepsilon_{max}$ (right axes) along the [100]-direction for (**a**) the ZFC, (**b**) the positively poled, and (**c**) the negatively poled states, at $T$=27 °C and with the frequency of 20 kHz. Here, the $\varepsilon_{max}$ is a maximum value of the dielectric constant of the ZFC state. (**d**) Successive 20 cycles' switching of $\varepsilon(E_{DC}=0)$ by applying electric field pulses of $E_{DC}=\pm 68$ kV/cm to the positively (red) and the negatively (blue) poled states. Grey, open-red, and open-blue arrows depict $P_{FE}$, positively poled $P_d$, and negatively poled $P_d$, respectively. The antiparallel and parallel polar-ordering between $P_{FE}$ and $P_d$ give rise to high and low dielectric states, respectively. As the switchings are repeated, the discrepancy between dielectric constants $\Delta\varepsilon(E_{DC}=0)$ of the high and low dielectric states is relaxed and reaches $\Delta\varepsilon$~30 (2.3 %), which is estimated by a relaxation function of $A\exp(-t/\tau)+\varepsilon$. $A$, $\tau$, and $\varepsilon$ for the high(low) dielectric state are $A$~25 (-12), $\tau$~198 (83) seconds, and $\varepsilon$~1330 (1300), respectively. Time dependence of applied $E_{DC}$ for the successive switching experiment of $\varepsilon(E_{DC}=0)$ in (**d**). The $E_{DC}$ was turned on for 2 seconds and off for 10

seconds. **e**, Three types of in-plane polar states of BaTiO$_3$ film on BaZrO$_3$ substrate and switchable high/low dielectric states of the in-plane polar states.

# Supplementary Materials

Switchable ferroelectric bias promoted by boosting square-tensile-strain

Jun Han Lee[1], Nguyen Xuan Duong[2], Min-Hyoung Jung[3], Hyun-Jae Lee[4], Ahyoung Kim[5], Youngki Yeo[6], Junhyung Kim[7], Gye-Hyeon Kim[1], Byeong-Gwan Cho[8], Jaegyu Kim[6], Furqan Ul Hassan Naqvi[9], Jeehoon Kim[10], Chang Won Ahn[2], Young-Min Kim[3], Tae Kwon Song[11], Jae-Hyeon Ko[9], Tae-Yeong Koo[8], Changhee Sohn[1], Kibog Park[1,7], Chan-Ho Yang[6], Sang Mo Yang[5], Jun Hee Lee[4], Hu Young Jeong[12], Tae Heon Kim[2,*], and Yoon Seok Oh[1,*]

*Correspondence to: thkim79@ulsan.ac.kr; ysoh@unist.ac.kr

DFT calculations.

We performed first-principles calculations based on density-functional theory (DFT) using the VASP code[1]. We adopted a local density approximation (LDA)[2] to describe the exchange correlation energy functional and pseudopotentials generated under a projector-augmented planewave (PAW) scheme[3]. The energy cut-off for the plane wave basis was set to 500 eV, and the force tolerance for the structure optimization was 0.001 eV/Å. For the unit cell of $BaTiO_3$ containing 5 atoms, a 6×6×6 $k$-point grid was chosen under the Monkhorst-Pack (MP) method[4] for sampling integrations over the Brillouin zone. Based on these conditions, our calculated lattice constants for cubic $BaTiO_3$, $a=b=c=3.9507(9)$ Å, were consistent with theoretical values[5]. The electric polarization was calculated from the effective charges, which were determined by the Berry phase method[6].

In order to understand the microscopic structure of the defect-dipole resulting from an oxygen vacancy ($V_O$), we calculated 3×3×3 supercell structures containing a total of 27 formula units with/without a $V_O$. The 2×2×2 $k$-points grid was used for the $c/a<1$ tetragonal supercell. For the theoretical analysis of the tensile strain effect, we refered the calculated cubic lattice constant[7]. Thus, we assumed that $\eta_a$=+1.25 % and $\eta_c$=-0.50 % are applied along the in-plane and out-of-plane directions, respectively. So, the lattice constants of the supercell were set as $a=b=12.0491(9)$ Å and $c=11.7172(5)$ Å. We considered three cases in which an oxygen was removed from an equatorial site in the [100]-axis (EQS-a), an equatorial site in the [010]-axis

(EQS-b), and an apical site in the [001]-axis (APS) in the oxygen vertexes of the $TiO_6$ octahedron.

For the calculation of the $V_O$ effects in the centrosymmetric phase, we made a $V_O$ from the tetragonal centrosymmetric (non-polar) phase. There are two sites of EQS-a/EQS-b and APS for $V_O$ formation (Fig. S1a), which have different symmetries in the centrosymmetric structure. In both cases, a $V_O$ leads to antiparallel displacement of two $Ti^{3+}$ ions (Fig. S1b and S1c), where the net electric dipole is zero. Energetically, $\Delta E$(EQS-a/EQS-b)-$\Delta E$(APS)=-119.7 meV; thus, $V_O$ formation at EQS-a/EQS-b is favored over that at APS.

For the $V_O$ effect in the polar phase, we induced the $V_O$ within the polar phase. The direction of $P_{FE}$ could be either [100]- or [010]-direction because of the 4-fold symmetry in the *ab*-plane (in-plane of the $BaTiO_3$ film). Herein, we assumed that the $P_{FE}$ was oriented along the [100]-direction in the polar phase. Then, we carried out calculations for Vo formation at EQS-a, EQS-b, and APS in the polar phase. In all three cases, $V_O$ produces a finite net electric dipole moment along the opposite direction ([$\bar{1}$00]-direction) of $P_{FE}$. Remarkably, the $V_O$ formation at EQS-b in the polar phase brings the lowest total energy ($E$=-1181.4052 eV), where the canted displacement of $Ti^{3+}$ ions leads to the electric dipole moment $P_d$ (Fig. S1e). It reveals that the canted $Ti^{3+}$-$V_O$-$Ti^{3+}$ complex is an essential unit to produce the $P_d$. The total energies of $V_O$ formation at EQS-a and APS in the polar phase are +46 meV and +128.2 meV higher than that of $V_O$ formation at EQS-b in the polar phase. Total electric polarization $P_{VO}$ for the $V_O$ formation at EQS-b is 7.5 μC/cm². Thus, $P_d$=$P_{VO}$-$P_{FE}$(=27.5 μC/cm²)=-20.0 μC/cm² for the 3×3×3 supercell. The electric polarization $P_{cmplx}$ of the $Ti^{3+}$-$V_O$-$Ti^{3+}$ complex, consisting of two unit cells, is $P_{cmplx}$=[(27×$P_{VO}$)-(25×$P_{FE}$)]/2=-242.5 μC/cm². In the same manner, ($P_{VO}$, $P_{cmplx}$) for the $V_O$ formations at EQS-a and APS are (22.4 μC/cm², -41.4 μC/cm²) and (1.7 μC/cm², -321.5 μC/cm²), respectively.

Interdigital electrodes.

The interdigital electrode pattern was formed by using ultraviolet (UV) photolithography processes. The interdigital electrode has 50 pairs of fingers with a finger length of 170 μm, a width of 9 μm, and a gap spacing of 5 μm. Due to the significant aspect ratio of the distance to the film thickness (~66 nm), we calculated the electric field by dividing the applied voltage by the distance. The AZ5214E photoresist film was spin-coated at 4000 rpm for 30 sec onto an HMDS (AZ AD Promoter-K) pre-treated $BaTiO_3$ surface, and the sample was baked on a hot plate at 105 °C for 120 sec. UV light of 130 mJ/cm$^2$ was exposed onto the photoresist-coated sample through a photomask containing the interdigital pattern by using a mask aligner (MA6, SUSS MicroTec, Germany). Then, the photoresist film was developed with AZ 300 MIF for 60 sec and rinsed subsequently in deionized water to remove the UV-exposed area. After the development process, a Ti/Au (10/50 nm) metal film stack was deposited onto the sample surface by using an e-beam evaporator (FC-2000, Temescal, USA). Finally, the sample was immersed in acetone for 12 hr to carry out the lift-off process to remove the remaining photoresist film together with the metal film stack on top of it and leave the interdigital electrodes in contact with the BaTiO$_3$ surface.

*P(E) measurements.*

The electric polarization vs. electric field hysteresis loops $P(E)$ were measured by homemade Sawyer-Tower circuits. All $P(E)$ measurements were performed in a vacuum chamber of Quantum Design PPMS to avoid dielectric break-down through the air gap or the specimen between the interdigital electrode's fingers. The upper limit of temperature control of the PPMS is $T=128$ °C so that we could heat up to $T=128$ °C. The electric fields at a frequency of 5 kHz for $E//[110]$ and $E//[100]$ were applied using a programmable function generator (HP33120) and a high voltage amplifier, and the current vs. $E$ data were recorded using an oscilloscope (TDS754) and a low noise current preamplifier (SR570). Polarization was achieved by integrating the current density $J$ vs. $E$ data. The electric field for the poling process was applied using a Keithley 2400 source meter. We gradually increased $E_{pole}$ to find an optimal magnitude of $E_{pole}$ without the dielectric break-down. $|E_{pole}|\sim 45$ kV/cm is large enough to pole $P_d$ along [100] and $|E_{pole}|\sim 68$ kV/cm for the [110] direction at $T=120$ °C (Fig. S13).

XRD and RSM measurements.

XRD and reciprocal space mapping were performed by using a four-circle high-resolution X-ray diffractometer (D8 Discover, Bruker). All the films were aligned with the known lattice constant of the BaZrO$_3$ substrate. For RSM, the Lynxeye 1D detector was utilized.

Raman Spectroscopy measurements.

Raman spectra were recorded at room temperature with a WITec Alpha 300R confocal Raman microscope system, equipped with a ×50 objective lens (numerical aperture = 0.8). Nd:YAG diode-pumped solid-state laser was employed for the excitation line of 532 nm, which was calibrated with a standard silicon wafer for each measurement. The power of the laser was 10 mW for all spectra. The spectrometer was an ultra-high throughput WITec UHTS 300 equipped with 1800 g mm$^{-1}$ grating.

Physical property measurements.

The temperature dependence of the magnetization was measured using the vibrating sample magnetometer (VSM) option in the Quantum Design PPMS-14T. Temperature dependent measurements of the dielectric-constant ($\varepsilon$) of the BaZrO$_3$ single-crystal were performed with oscillating electric fields applied along the [100] direction using a Hewlett-Packard HP4275A LCR meter at 20 kHz. Optical transmittance spectra were measured by an Agilent Cary 5000. The DC electric field dependence of dielectric constant was measured using a Hewlett-Packard HP4275A LCR meter and Keithley 2400 source meter.

Temperature-dependent XRD measurements.

XRD measurements as a function of the temperature were performed at the beamline (3A) of PLS-II. We conducted $\theta-2\theta$ scans to investigate the structural variations in the in-plane and out-of-plane lattice parameters of the BaTiO$_3$ film and BaZrO$_3$ substrate. The lattice parameters were traced using a closed-loop helium gas refrigerator with a high-temperature interface in the range from 50 °C to 460 °C.

TEM measurements.

Cross-sectional samples of the heteroepitaxial BaTiO$_3$ film grown on (001) BaZrO$_3$ substrate were prepared by the Ga ion beam milling method using a dual-beam focused ion beam system (FIB, Helios NanoLab 450, Thermo Fisher Scientific). For reducing the sample damage from Ga-implantation of FIB, the samples were additionally milled by using a low-energy Ar-ion milling system (Fischione model 1040 Nanomill). STEM images were obtained using a FEI Titan G2 60-300 with a double Cs corrector at an accelerating voltage of 200kV.

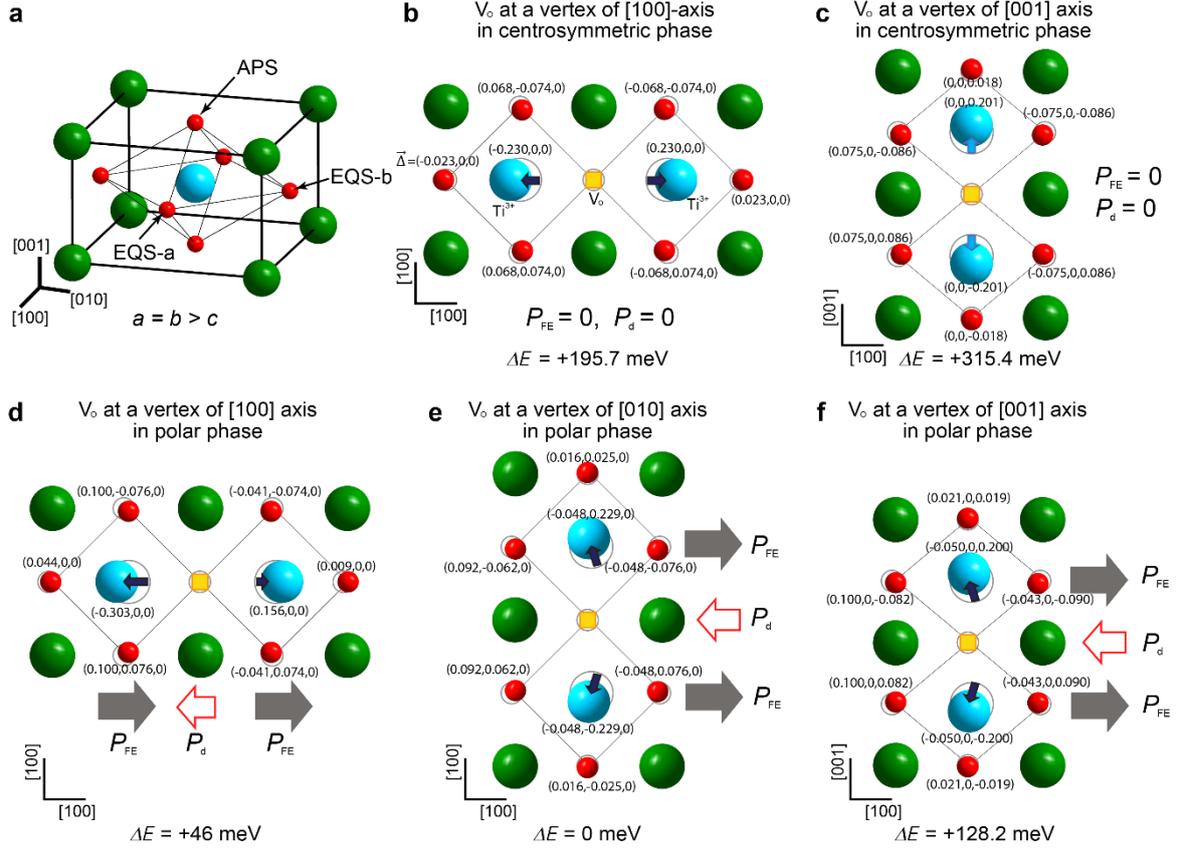

**Fig. S1. Schematics of the local displacement of $Ti^{3+}$ and $O^{2-}$ ions for $V_O$ formation. a**, Schematic of a *c/a*<1 tetragonal unit cell of square-tensile-strained BaTiO$_3$. The green, blue, and red spheres represent barium ($Ba^{2+}$), trivalent titanium ($Ti^{3+}$), and oxygen ($O^{2-}$) ions, respectively. (**b-f**) Local structure in the *ab*-plane or the *ca*-plane, simulated by density functional theory (DFT) calculations of $Ti^{3+}$ and $O^{2-}$ ion displacements for a $V_O$ in (**b-c**) the centrosymmetric tetragonal phase and (**d-f**) the polar phase, where $P_{FE}$ is oriented along the [100]-direction, in the 3×3×3 supercell. The $V_O$ is denoted as yellow squares. $\vec{\Delta}$ depicts the relative displacement of $Ti^{3+}$ and $O^{2-}$ ions from their initial position without $V_O$. *ΔE* indicates the relative total energy to the total energy (**e**) when a $V_O$ is formed at a vertex of the [010]-axis in the polar phase (*E*=-1181.4052 eV). (**b-c**) A $V_O$ at a vertex of (**b**) the [100]- and (**c**) the [001]-axis leads to antiparallel displacement of $Ti^{3+}$ ions, where the net electric dipole is zero. Energetically, the antiparallel displacement of $Ti^{3+}$ ions for $V_O$ at the vertex of the [100]-axis yields -119.7 meV lower total energy than that of $V_O$ at the vertex of the [001]-axis. (**d-f**) $Ti^{3+}$ and $O^{2-}$ ion displacements for $V_O$ at a vertex of (**d**) the [100]-, (**e**) the [010]-, and (**f**) the [001]-axis in the polar phase, where $P_{FE}$ is along the [100]-direction. The DFT calculations for the polar phase along the [100]-direction without $V_O$, thus $V_O$-less BaTiO$_3$, result in $P_{FE}$=27.5 μC/cm$^2$. **d**, The $V_O$ at the vertex of the [100]-axis produces a defect-dipole moment with asymmetric displacement of $Ti^{3+}$ ions, $P_d = -5.0$ μC/cm². In contrast, $V_O$ at a vertex of (**e**) the [010]- and (**f**) the [001]-axis in the polar phase develops symmetrically canted displacement of $Ti^{3+}$ ions and finite defect-dipole moments of $P_d$=-20.0 μC/cm² and $P_d$=-25.8 μC/cm², respectively, along the opposite direction of the $P_{FE}$. Among these three $V_O$ formations in the polar phase, (**e**) the $V_O$ at the vertex of the [010]-axis has the lowest total energy.

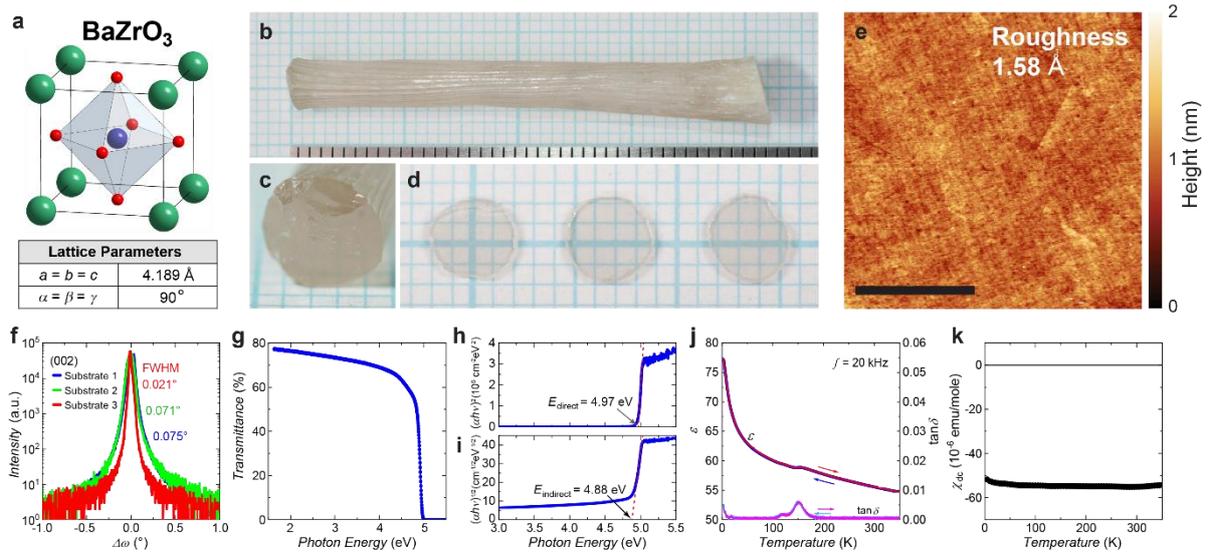

**Fig. S2. 4.189 Å cubic perovskite BaZrO$_3$ single-crystal substrate and physical properties. a**, Schematic of a unit cell of cubic perovskite BaZrO$_3$ ($Pm\bar{3}m$ space group) and the corresponding lattice parameters. The green, blue, and red spheres represent barium (Ba), zirconium (Zr), and oxygen (O), respectively. Photographic images of (**b**) a single-crystalline BaZrO$_3$ rod grown by the optical floating zone method, (**c**) a cleaved (001) surface of the rod, and (**d**) sectioned/polished BaZrO$_3$ substrates of the (001) surface. **e**, Atomic force microscopy (AFM) image of the polished (001) surface of the BaZrO$_3$ single-crystal substrate. The root-mean-square (RMS) surface roughness is as low as 1.58 Å. The scale bar depicts 4 μm. **f**, Rocking curves of the (002) Bragg peak of the BaZrO$_3$ substrates. The full width half maximum (FWHM) is 0.021–0.075°. **g**, Optical transmittance spectra and Tauc plots $(\alpha \cdot h\nu)^{1/r}$ with exponent $r$ values of (**h**) 1/2 and (**i**) 2. Here, $\alpha$, $h$, and $\nu$ are the absorption coefficient, Planck constant, and frequency, respectively. The red dashed lines in (**h**) and (**i**) depict linear fitting lines at the band edge to determine the direct and indirect bandgaps. Direct and indirect bandgaps of (**h**) 4.97 eV and (**i**) 4.88 eV, respectively, are estimated for our BaZrO$_3$ single-crystal substrate. Temperature dependence of the (**j**) dielectric constant, loss tangent delta, and (**k**) magnetic susceptibility of the BaZrO$_3$ substrate.

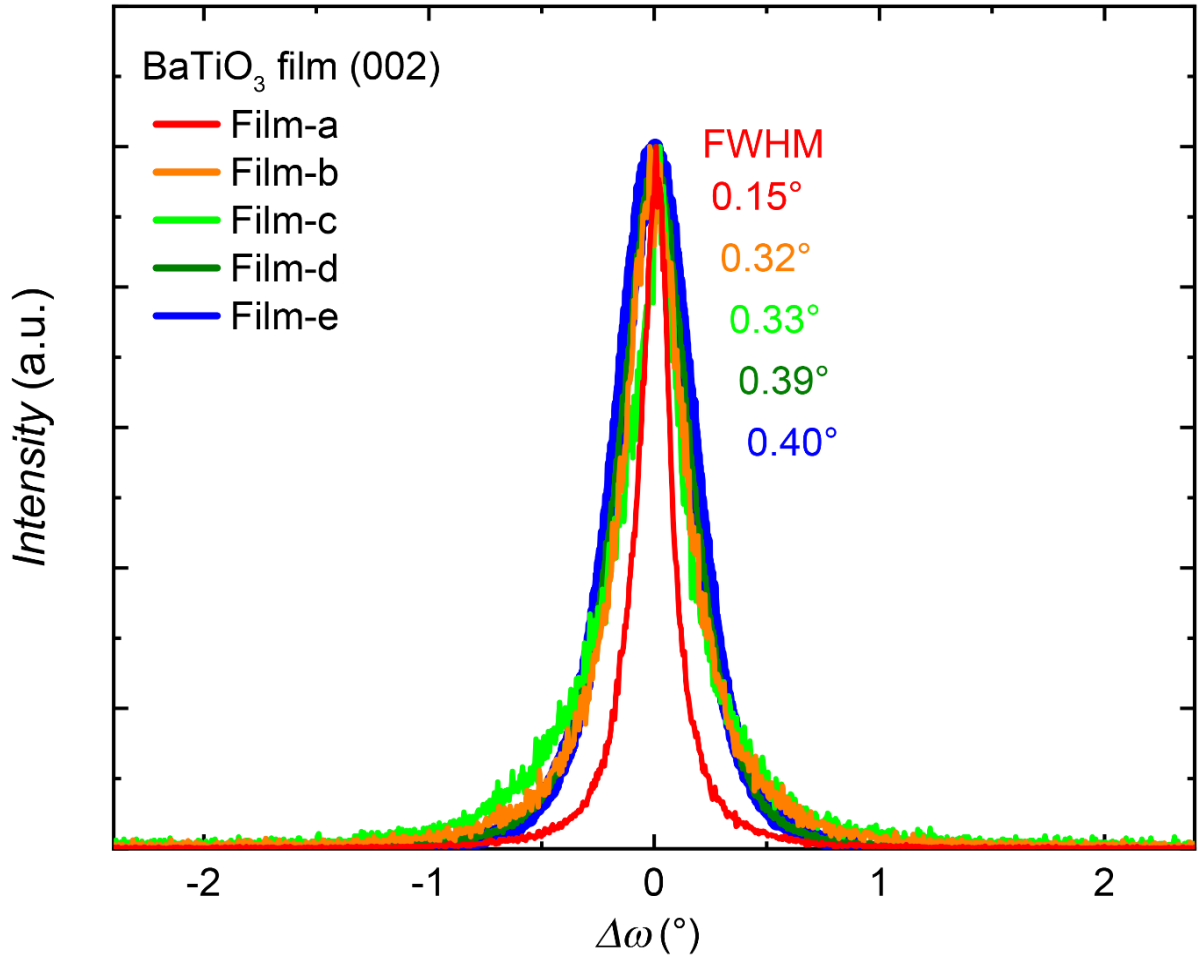

**Fig. S3. Rocking curves of the epitaxially BaTiO₃ films grown on the BaZrO₃ substrates.** Rocking curves of the (002) Bragg peak of the BaTiO₃ films on the BaZrO₃ substrates. The full width half maximum (FWHM) values of film-a (red), b (orange), c (green), d (dark green), and e (blue) are 0.15, 0.32, 0.33, 0.39 and 0.40°, respectively.

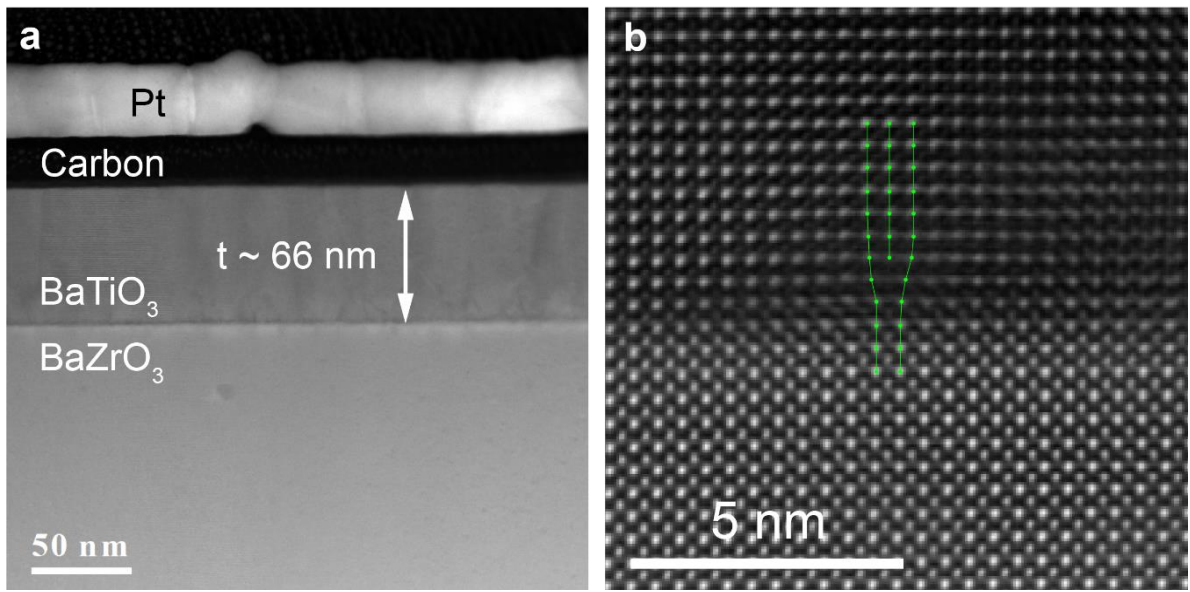

**Fig. S4. Cross-sectional transmission electron microscopy (TEM) image. a**, Cross-sectional annular dark-field (ADF) scanning transmission electron microscopy (STEM) image of $BaTiO_3$/$BaZrO_3$ with carbon and platinum (Pt) coating. The thickness of $BaTiO_3$ film is determined to be 66 nm. **b**, Edge dislocation at the interface between the $BaTiO_3$ film and the $BaZrO_3$ substrate. The green points and lines guide Ba-Ba unit cell edges.

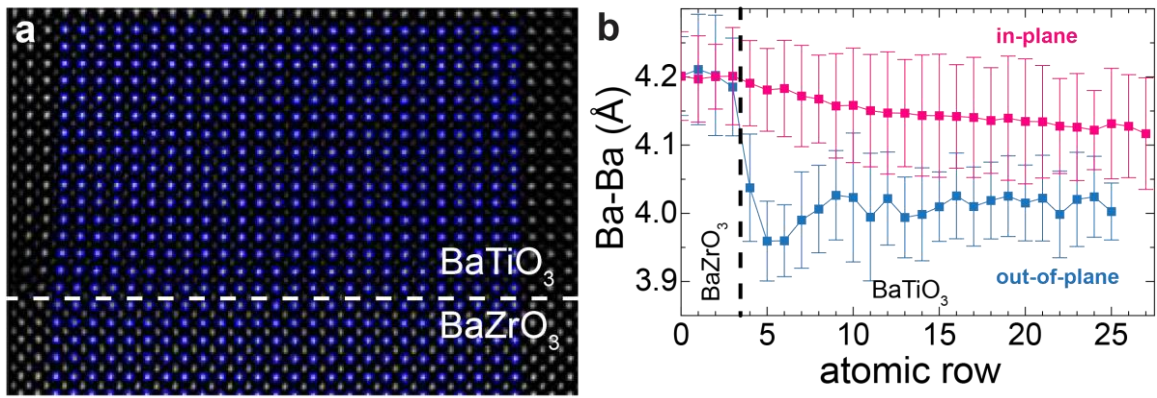

**Fig. S5. Atomic-scale resolution annular dark field (ADF) scanning transmission electron microscope (STEM) image at the interface of BaTiO₃ film and BaZrO₃ substrate. a,** Atomic-scale resolution of cross-sectional ADF-STEM image at the BaTiO₃/BaZrO₃ interface. **b,** Variation of in-plane and out-of-plane Ba-Ba ion spacing near the interface. The colored map in (**a**) and in-plane ion spacing in (**b**) show that in-plane lattice constants partially relax away from the interface.

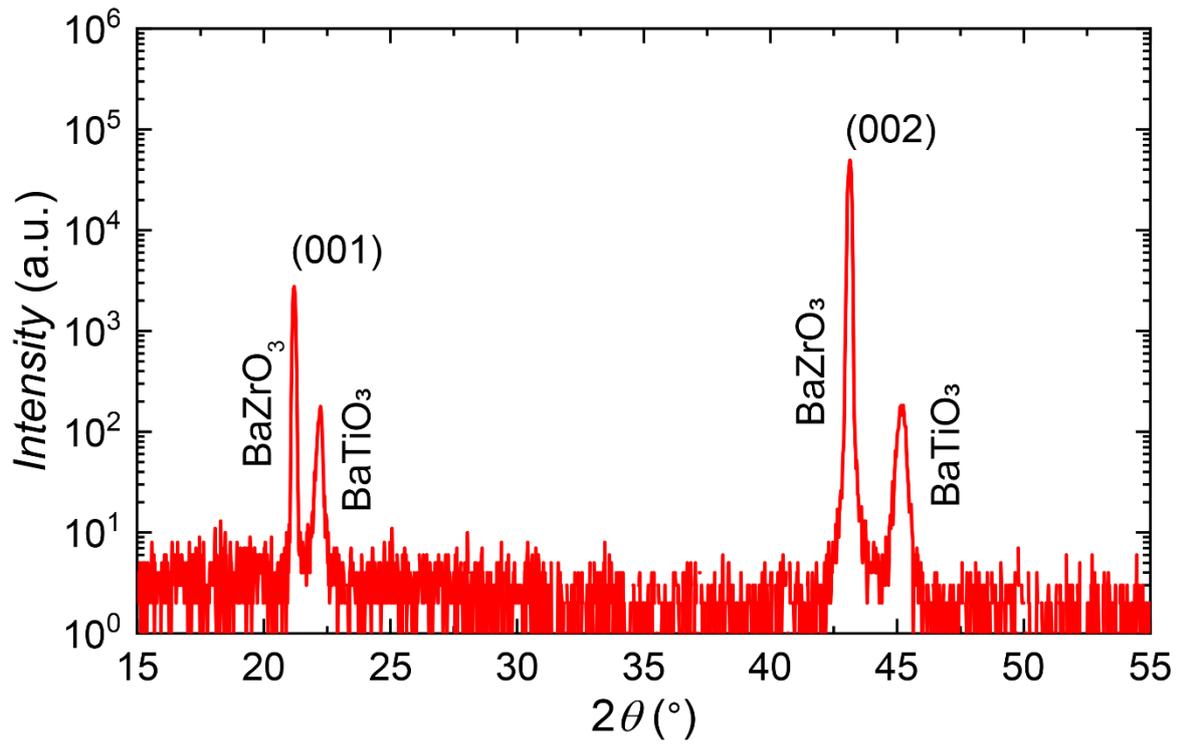

**Fig. S6.** *θ-2θ* x-ray diffraction pattern of an epitaxial BaTiO$_3$ film grown on a (001) BaZrO$_3$ substrate.

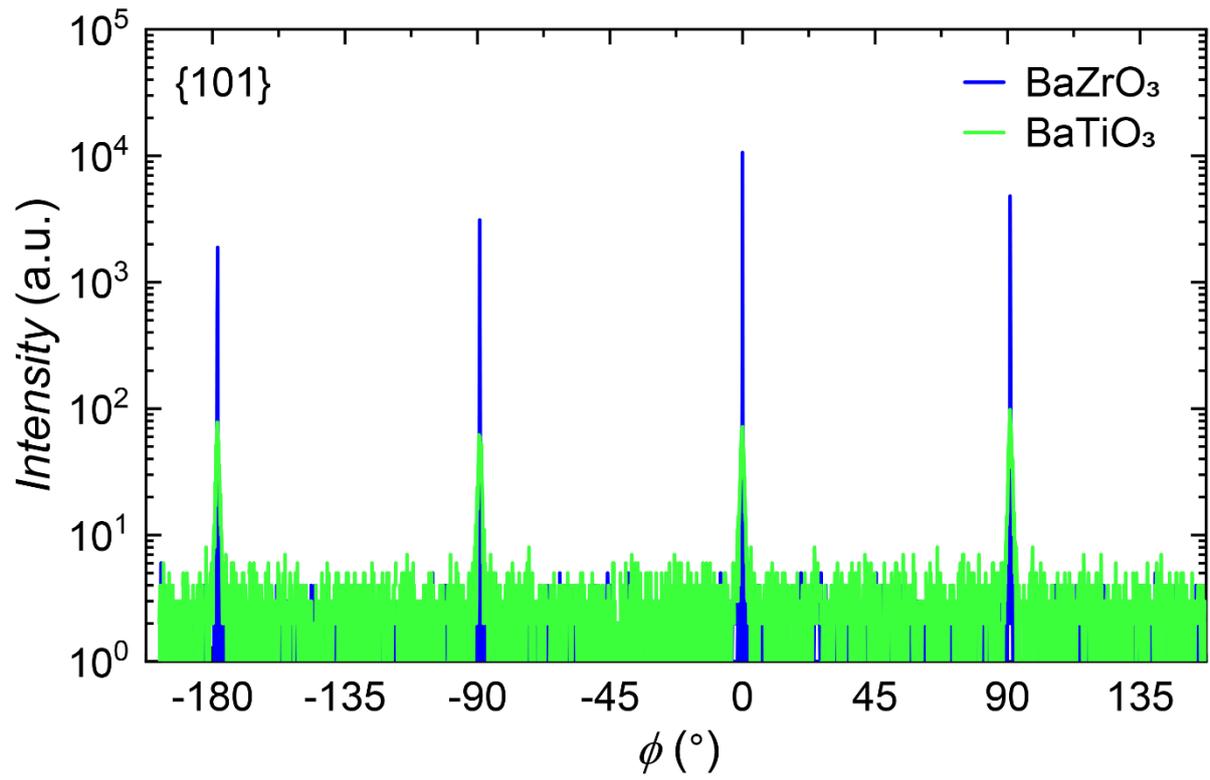

Fig. S7. $\phi$-scan of the (101) Bragg peaks of the BaTiO$_3$ film (green) on the BaZrO$_3$ substrate (blue).

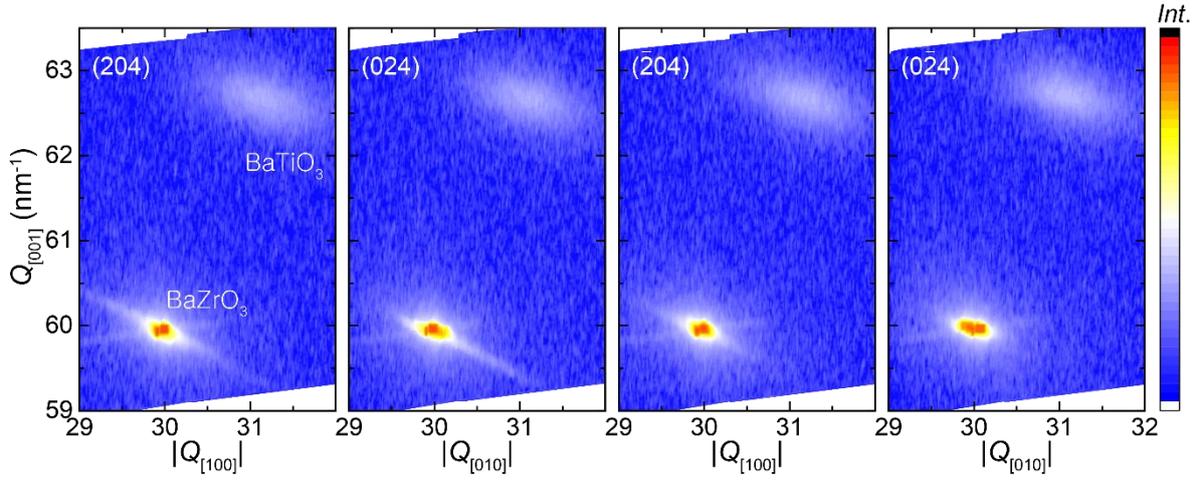

**Fig. S8. Reciprocal space maps (RSMs) at (204) reflections for the BaTiO$_3$ film and BaZrO$_3$ substrate.** The z direction corresponds to out of the sample surface plane. Strong and sharp BaZrO$_3$ substrate peaks exist near $Q_{[001]}$=59.96 nm$^{-1}$ and $|Q_{[100]}|=|Q_{[010]}|$=29.98 nm$^{-1}$, and relatively weak BaTiO$_3$ film peaks are observed near $Q_{[001]}$=62.68 nm$^{-1}$ and $|Q_{[100]}|=|Q_{[010]}|$=31.11 nm$^{-1}$., which indicate $c$=4.009(7) Å and $a$=$b$=4.039(3) Å, respectively. This indicates that, with respect to the pseudo-cubic lattice constant $a_{pc}$=4.000(4) Å ($a_{pc}=V^{1/3}$; unit cell volum $V$=63.97 Å$^3$) of the bulk BaTiO$_3$ at $T$=20 °C[8], in-plane tensile strain 0.98 % is applied to the BaTiO$_3$ film. The BaTiO$_3$ film (unit cell volume $V$=65.40 Å$^3$) on the BaZrO$_3$ substrate has a 2.24 % unit cell volume expansion with respect to bulk tetragonal BaTiO$_3$ at $T$=20 °C. The intensity is described with a log scale, shown in the colour bar.

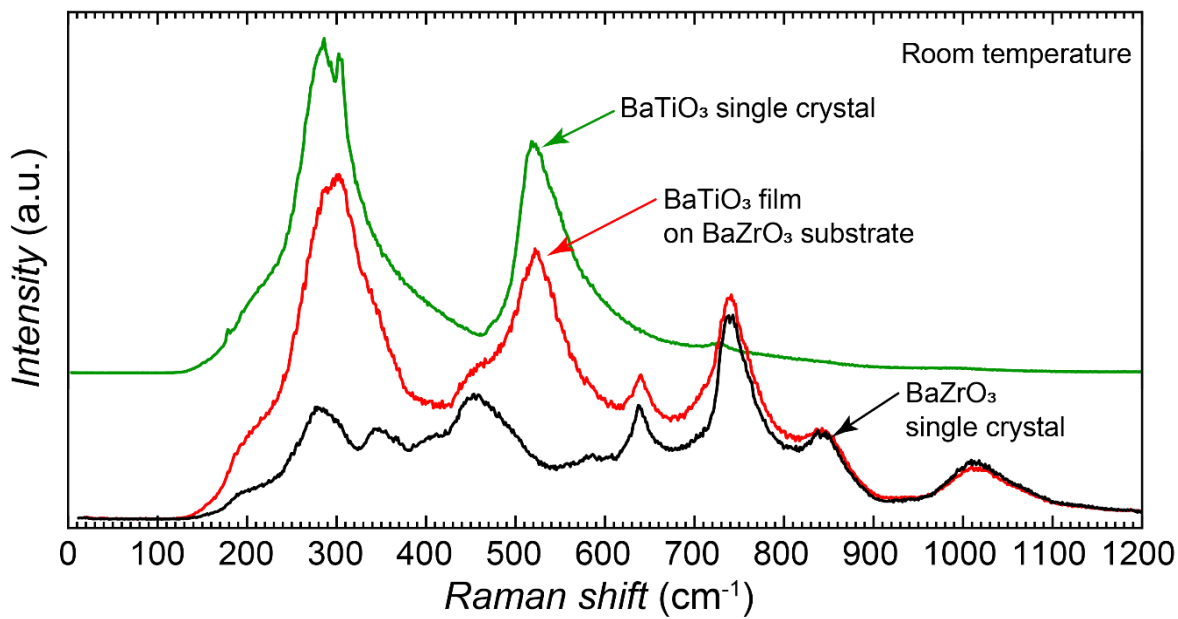

**Fig. S9. Raman spectra.** Raman spectra of the BaTiO$_3$ single-crystal (green), the BaTiO$_3$ film on the BaZrO$_3$ substrate (red), and the BaZrO$_3$ single-crystal (black) at room temperature.

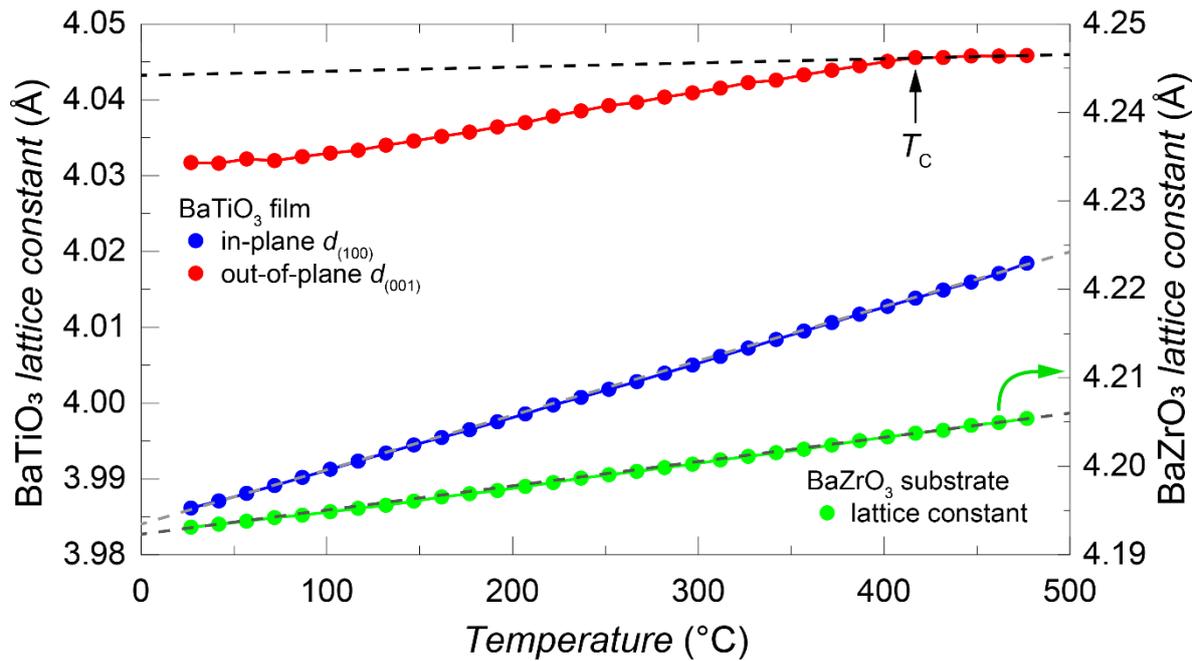

**Fig. S10. Temperature dependence of lattice constants of the BaTiO$_3$ film and the BaZrO$_3$ substrate.** The in-plane $d_{(100)}$ (red) and out-of-plane $d_{(001)}$ (blue) lattice constants of the BaTiO$_3$ thin film were estimated from the (002) and (201) Bragg reflections of the x-ray diffraction. The lattice constant of the BaZrO$_3$ substrate (green) was obtained from the (002) Bragg reflections. The dashed lines guide the linear temperature dependence of the lattice constants. The temperature dependence of the in-plane lattice constants of the BaTiO$_3$ film exhibits an anomaly at $T\sim415$ °C, but the out-of-plane lattice constant linearly increases without an anomaly up to $T=475$ °C. The change in slope at $T\sim415$ °C reflects a phase transition of the BaTiO$_3$ film.

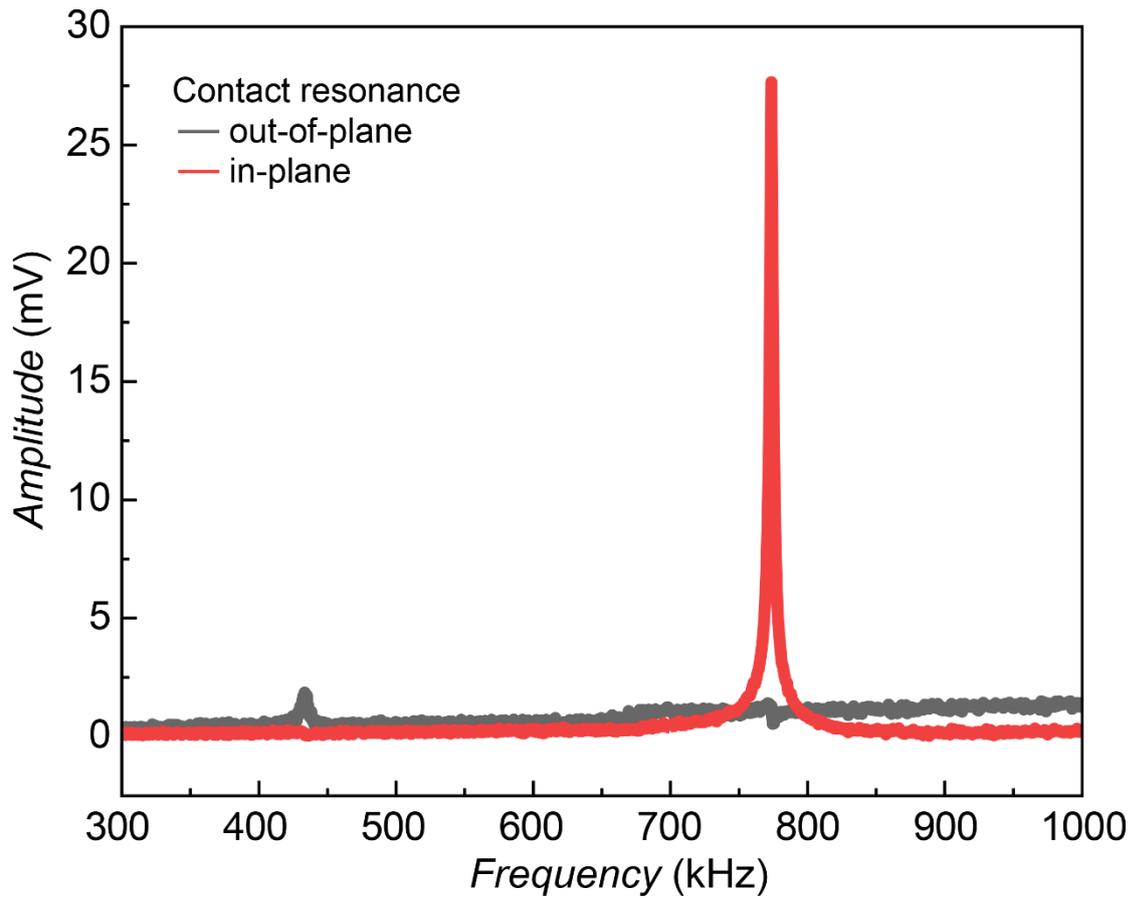

**Fig. S11. Contact resonance of the piezo-response force microscopy (PFM) signals.** Contact resonance of the piezo-response force microscopy (PFM) signals for the in-plane (red) and out-of-plane (grey) directions while applying the AC voltage along the [110] direction of the interdigital electrodes. The contact resonance appears at *f*=770 kHz for the in-plane and *f*=430 kHz for the out-of-plane direction. A predominant in-plane piezo-response is observed.

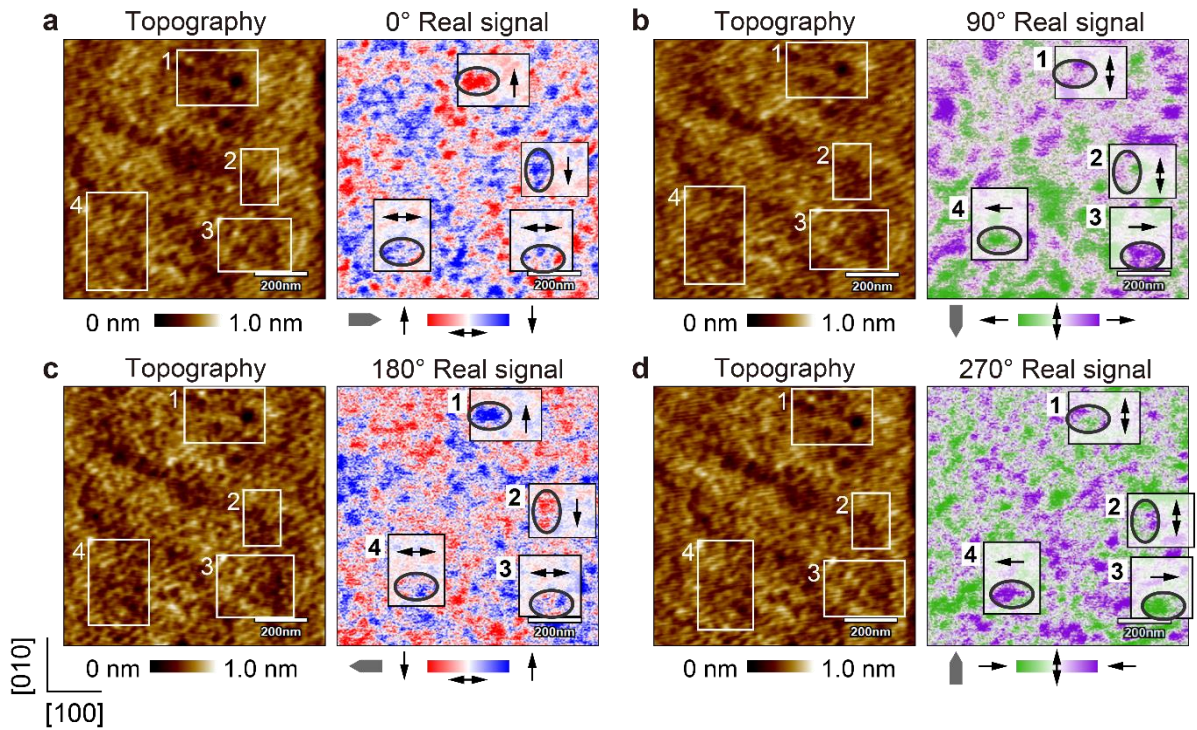

**Fig. S12. Observation of an in-plane PFM domain structure in the BaTiO₃ film on the BaZrO₃ substrate.** Topographic image and in-plane PFM image simultaneously taken by a tip of which the cantilever axis orients along (**a**) 0°, (**b**) 90°, (**c**) 180°, and (**d**) 270°. The cantilever axis points to the [100]-direction in the geometry of 0° orientation. Tip orientation and the expected piezoresponse directions are drawn by tip cartoon on each of images.

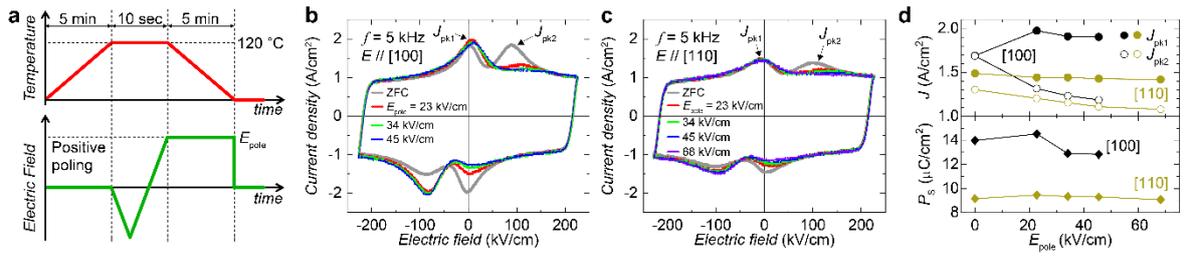

**Fig. S13. Switching current density versus electric field of the *P(E)* measurements and the saturated electric polarization for various poling electric fields. a,** Poling process for the positive poling. The specimen was warmed up to *T*=120 °C. Applying a constant electric field $E_{\text{pole}}$ and following a three-quarter wave along the [100] or [110] direction with *f*=0.1 Hz at *T*=120 °C, the specimen was cooled down from *T*=120 °C to room temperature (*T*=27 °C) under the applied $E_{\text{pole}}$. After turning off the $E_{\text{pole}}$ at room temperature, we measured *P(E)*. The negative poling follows the same process but with the opposite sign of $E_{\text{pole}}$. **b,c,** Current density *J* versus electric field when the electric field *E* is applied along (**b**) the [100] and (**c**) [110] directions after the poling process with various poling electric fields $E_{\text{pole}}$ and zero-field cooling (ZFC, grey). $J_{\text{pk1}}$ and $J_{\text{pk2}}$ indicate *J* values at the peak positions while *E* ramps up forward to positive *E* during the *P(E)* measurements. **d**, $E_{\text{pole}}$ dependence of $J_{\text{pk1}}$ (closed circles), $J_{\text{pk2}}$ (open circles), and saturated polarization $P_S$ (closed diamonds) for the [100] (black) and [110] (yellow) directions. Application of $E_{\text{pole}}$=23 kV/cm yields distinct suppression of the $J_{\text{pk2}}$ for both [100] and [110] directions and enhancement of $J_{\text{pk1}}$ and $P_s$ for the [100] direction. As increasing $E_{\text{pole}}$, $J_{\text{pk2}}$ keep decreasing, but the decrement becomes slow down. For the thermal poling process on the $BaTiO_3$ film, $E_{\text{pole}}$=45 kV/cm and 68 kV/cm were applied for [100] and [110] directions, respectively.

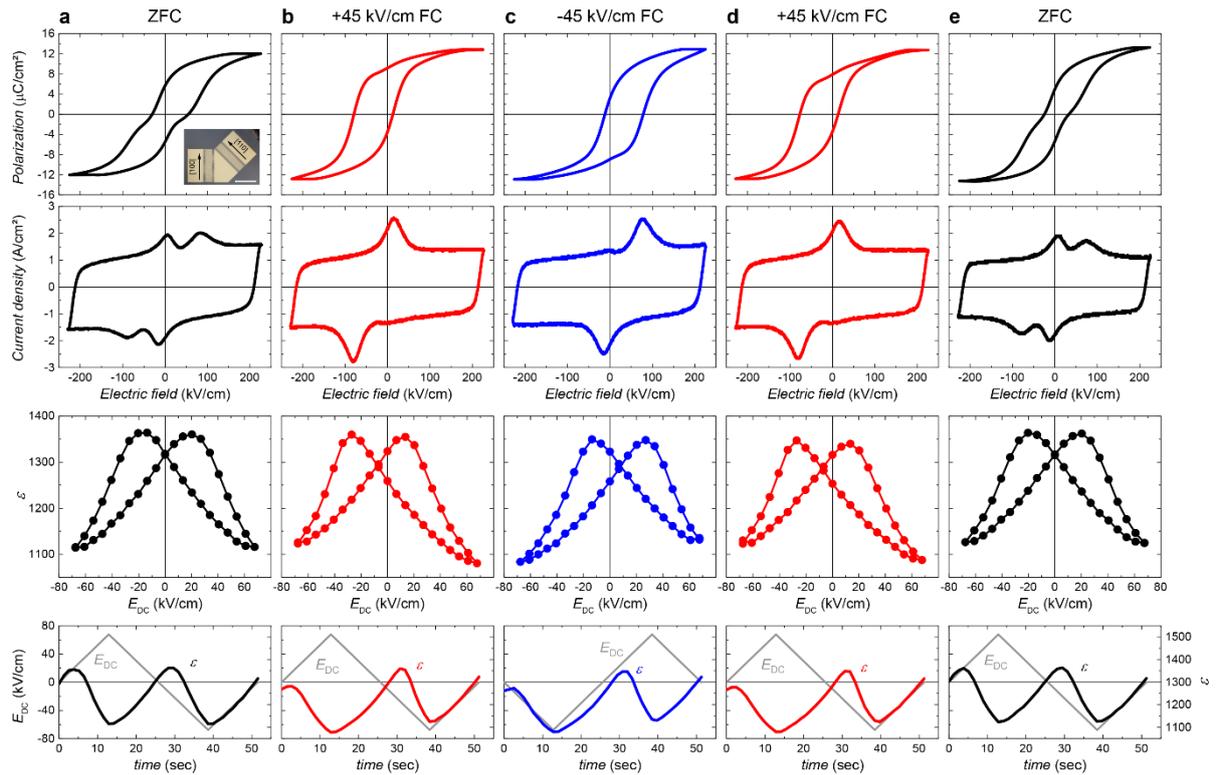

**Fig. S14. Successive switching of the three-polar ordering states through the thermal poling process.** $P(E)$ hysteresis loop, switching current density versus electric field of the $P(E)$ measurement, dielectric constant $\varepsilon$ versus DC electric field $E_{DC}$, and time dependent $E_{DC}$ profile of $\varepsilon(E_{DC})$ measurement after each successive polings of (**a**) zero-field-cooling(ZFC), (**b**) $E_{pole}$=+45 kV/cm poling, (**c**) $E_{pole}$=-45 kV/cm poling, (**d**) $E_{pole}$=+45 kV/cm poling, and (**e**) ZFC again. The inset of (**a**) depicts a photo of the interdigital electrodes on the film